# Immersive Human-Machine Teleoperation Framework for Precision Agriculture: Integrating UAV-based Digital Mapping and Virtual Reality Control

Tao Liu[a], Baohua Zhang[a*], Qianqiu Tan[a], Jun Zhou[c], Shuwan Yu[a], Qingzhen Zhu[b], Yifan Bian[a]

[a]College of Artificial Intelligence, Nanjing Agricultural University, Nanjing, Jiangsu, PR China

[b]School of Agricultural Equipment Engineering, Jiangsu University, Zhenjiang, Jiangsu, PR China

[c]College of Engineering, Nanjing Agricultural University, Nanjing, Jiangsu, PR China

*Correspondence should be addressed to Baohua Zhang (bhzhang@njau.edu.cn). The primary contact's telephone number is +86 (025) 5860 6585 and the fax number is +86 (025) 5860 6585.

**ABSTRACT**

In agricultural settings, the unstructured nature of certain production environments, along with the high complexity and inherent risks of production tasks, poses significant challenges to achieving full automation and effective on-site machine control. Remote control technology, which leverages human intelligence and precise machine movements, ensures operator safety and boosts productivity. Recently, virtual reality (VR) has shown promise in remote control applications by overcoming single-view limitations and providing three-dimensional information, yet most studies have not focused on agricultural settings. Therefore, to bridge the gap, this study proposes a large-scale digital mapping and immersive human-machine teleoperation framework specifically designed for precision agriculture. In this research, a DJI unmanned aerial vehicle (UAV) was utilized for data collection, and a novel video segmentation approach based on feature points was introduced. To accommodate the variability of complex textures, this method proposes an enhanced Structure from Motion (SfM) approach. It integrates the open Multiple View Geometry (OpenMVG) framework with Local Features from Transformers (LoFTR). The enhanced SfM produces a point cloud map, which is further processed through Multi-View Stereo (MVS) to generate a complete map model. For control, a closed-loop system utilizing TCP/IP for VR control and positioning of agricultural machinery was introduced. This system offers a fully visual-based method for immersive control, allowing operators to utilize VR technology for remote operations. The experimental results demonstrate that the digital map reconstruction algorithm developed in this study offers superior detail reconstruction, along with enhanced robustness and convenience. The user-friendly remote control method also showcases its advantages over traditional video streaming-based remote operations, providing operators with a more comprehensive and immersive experience and a higher level of situational awareness.

## 1. Introduction

The advent of agricultural machinery has brought about a revolutionary change in agricultural production, with its widespread use significantly reducing the need for manual labor (Jin et al., 2021; Thakur et al., 2023). However, traditional agricultural machinery still requires on-site human operation, which not only increases the transportation costs for agricultural workers but also poses potential health risks to the operators. Emerging autonomous operation technologies have mitigated these issues to some extent. Nowadays, agriculture is moving towards the incorporation of modern machinery that is based on automation and robotics, aiming to improve production efficiency through the automatic operation of machinery (Thakur et al., 2023). Agricultural machines are used to solve planting (Jin et al., 2021), transplanting (Zhang et al., 2024), picking (Ji et al., 2022). Although the automation of agricultural machinery operations can address such issues, it may not be fully applicable due to the non-structured natural conditions in some countries and the complexity of certain agricultural activities. For instance, in the mountainous terraced areas of southern China, the agricultural landscape is intricate, and the presence of various unforeseen circumstances renders the application of automated agricultural machinery impractical. For tasks like orchard pesticide spraying, where trees have diverse growth forms and spraying locations vary, coupled with the potential harm of pesticides to human health, both mechanical automation and manual spraying are infeasible. In this context, remote operation technology has emerged, effectively combining human intelligent decision-making with machine motion, endowing agricultural robots with human intervention capabilities (Benos et al.,2020). It provides an effective solution for precision agriculture and serves as a supplement to autonomous operations. Remote operation technology enhances the versatility and operational capabilities of agricultural machinery across diverse environments.

Traditional teleoperation methods often involve placing cameras at the operation site to capture visuals. These pieces of information are then displayed on a computer screen, guiding the operation of the robots (Green et al., 2021; Qin et al., 2022; Ersahin & Sedef, 2015). However, purely video-based robotic teleoperation and scene exploration are limited by the coupling of the view directly to the area observed by the camera, which affects the level of immersion and situational awareness (Stotko et al., 2019). Additionally, the positioning of relevant scene entities presented in the video stream can become more complex. These conventional methods based on video streams have significant limitations, including a single, limited viewpoint an inability to provide operators with 3D information and so on, which may impede efficient operation. In order to overcome these constraints, Sun et al. (2016) and others developed a system using four fish-eye cameras to provide an omnidirectional view around the operated machine. The system enables synchronized remote visualization with a 360° perspective. Valiton (2021) explored the use of two cameras to offer different crucial viewing angles, aiding operation. Experimental validation found that the use of multiple cameras to change perspectives can be beneficial for the operator's work. However, this method is not suitable for the acquisition of large-scale information around the machine. With advancements in display technology, VR devices have begun to garner attention from both consumers and researchers as they address issues such as low resolution and high display costs. More and more researchers have started turning to VR technology (Stotko et al., 2019; Bazzano, F., 2016; Pérez et al., 2019). Bian et al. (2018) introduced a three-dimensional visual feedback subsystem, built on VR technology, designed to provide a 3D perception to the operator. This system delivered immersive visual feedback to the operator, enhancing their spatial awareness during operation; Lipton et al. (2017) and Su et al. (2022) incorporated the operator into a VR control room, where the visual scenes captured from various camera perspectives were projected onto the VR headset. Sun et al. (2020) advanced an innovative approach that integrates Mixed Reality (MR) with remote machine operation. This method involved the development of a sophisticated MR interface, which effectively amalgamates elements of the real and virtual environments. Numerous studies have demonstrated the potential of VR in teleoperation, showing that it provides more immersive information about the operation site compared to traditional displays (Opiyo et al., 2021; Sun et al., 2020; Bazzano, F., 2016). This study builds on these advancements by using VR technology to develop comprehensive virtual models of large-scale agricultural environments. The aim is to establish an immersive human-robot interaction methodology for teleoperating agricultural mobile robots within these simulated environments.

Significant advancements have been made in digital map reconstruction due to improvements in computational capacity, algorithms, and image acquisition techniques. Lv et al. (2017), Zhang et al. (2018), and Sun et al. (2019) have utilized Kinect cameras to capture RGB-D data, initiating 3D reconstruction efforts across various fields. However, the depth acquisition range of depth cameras is relatively limited, making it unsuitable for direct reconstruction of large outdoor agricultural scenarios. The data collection method employed in this study involves using drones for aerial photography. UAVs, due to their mobility, are ideal for both aerial and ground photography. Gonçalves et al. (2021), Berra & Peppa (2020), and Meinen & Robinson et al. (2020) have used UAV to capture large-scale outdoor scene data, followed by 3D map reconstruction using the SfM-MVS method.

It is worth noting that 3D reconstruction remains a significant research field, having developed numerous effective algorithms, including the OpenMVS algorithm, the latest NERF algorithm, COLMAP, among others. Nevertheless, the quality of 3D reconstruction is influenced by several factors, such as Weak Texture Scenes (Yang & Jiang, 2021) and the shooting angle (Gonçalves et al. 2021), among others. Especially on regions with poor texture, the reconstruction has been a challenging issue, which is significantly influenced by the extraction, description, and matching methods of feature points. In this regard, the SIFT method, commonly used in SfM, does not perform well, and an insufficient number of feature points will directly affect the reconstruction outcome (Hoshi et al.,2022). In order to address this issue, Yang & Jiang (2021) addressed regions with poor texture by covering the light source with a film pattern, "adding" texture to the original scene, thereby enhancing weak textures into rich textures for processing. Hoshi et al (2022) combined SuperPoint, SuperGlue, and LoFTR from deep learning, using SuperPoint + SuperGlue to process regular image areas and LoFTR to deal with regions with poor texture. These research achievements have brought new possibilities for digital map reconstruction.

In the domain of Human-Robot Interaction (HRI), merging the virtual environment with the control of machines in the real world is an important area of research. Virtual environments commonly leverage gaming engines, exemplified by the widespread use of powerful platforms such as Unity and Unreal Engine. Concurrently, the Robot Operating System (ROS) serves as a prevalent framework on the machinery side. The fusion of gaming engines and ROS facilitates the amalgamation of virtual reality interaction and machine control. Krupke et al. (2018) proposed a software toolkit employing Unity for designing interactive virtual and real environments, achieving bidirectional communication between Unity3D and ROSbridge. To develop an automatic delivery robot, Liu et al. (2020) mapped the ROS robot's virtual model in Unity to a real robot. This approach allows for a successful perception match between the real world where the robot is located and the virtual world. Mizuchi (2017) incorporated Virtual Reality (VR) and proposed a new software architecture. This study provides a virtual identity for users in the virtual world, allowing interaction with the virtual environment through VR devices (headsets, controllers). Furthermore, a bridging mechanism between Unity and ROS was developed to support more rapid interactions. Meanwhile, Roldán et al. (2019) developed a unique Virtual Reality interface using multiple ROS packages and Unity assets. The resulting interface, built on the combination of ROS and Unity, is capable of controlling multiple robots. Connecting the virtual world with machine control offers many benefits. Especially today, With the rapid advancement of computers and display technologies, VR devices have seen significant improvements, along with a drastic reduction in costs and increased usability. This has made the advantages of VR devices more pronounced, making it easier for operators to access three-dimensional information and providing them with a more immersive display experience (Slob et al.,2023). When the virtual world accurately represents the real world, the operator can view it from any perspective without affecting the actual environment. This method of observation provides the operator with more information than direct observation through video streams. To influence the real world, the operator simply needs to activate the control button, allowing direct interaction with the environment.

Rooted in the goal of achieving immersive remote control, this research proposes a framework for the digital reconstruction of large-scale agricultural scenarios and immersive teleoperation. A UAV is employed for data collection. To accommodate the reconstruction of large-scale outdoor agricultural maps, a video segmentation algorithm was developed to reduce data volume. Following this, to adapt to agricultural reconstruction, improved SfM algorithms and MVS are utilized for the process. The resulting digital agricultural map is integrated into a VR system for immersive viewing. Our system uses the TCP/IP to facilitate real-time control and location tracking of agricultural machinery in the VR setting. In this research, the position of agricultural machinery is updated in real-time through the VR head-mounted display, providing operators with powerful capabilities for gathering on-site information. This research not only provides a novel approach for the remote operation of agricultural robots but also makes a contribution to the field of three-dimensional reconstruction.

## 2. Materials and methods
### 2.1 Overall Framework

To achieve immersive teleoperation, a system with an overall workflow is proposed, as depicted in Figure 1. The system consists of data acquisition, digital map reconstruction,

VR visualization, and immersive closed-loop control. The drone captures video footage to collect scene data, which is then matched and segmented frame-by-frame using feature points. The segmented data undergoes reconstruction using a multi-view geometry (OpenMVG) algorithm that integrates LOFTR, followed by the construction of a complete digital agricultural map using OpenMVS. The agricultural map, rendered with Unity, is then displayed immersively in VR. Teleoperation information exchange is facilitated by a direct connection between VR and the ROS robot. Through message communication, control messages are sent from VR to the robot, and the robot's position is updated in real-time within VR, forming a closed-loop control framework.

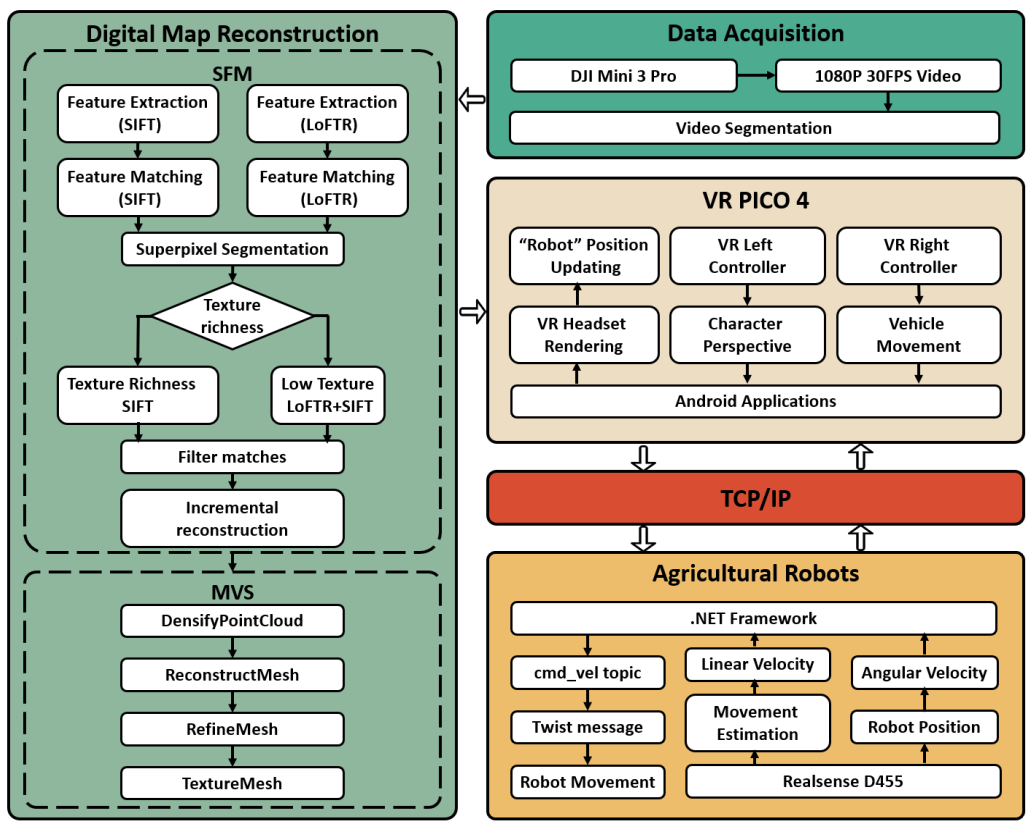

Figure 1 Overall Framework of the Proposed System

## 2.2 Study Site

Experiments were conducted in a total of four agricultural settings. The study sites were located in Ge Guan Ecological Park, Nanjing, China (32°12'N, 118°40'E). The park consists of three parts: a greenhouse garden, a greenhouse plantation, and an outdoor area. The control room is located inside the greenhouse garden. Study site 1 was an outdoor farmland cultivated with vegetables. Study site 2 was an uncultivated area. Both sites are illustrated in Figure 2b, which serves as a representative of the planar digital maps. Study site 3, which was part of the Greenhouse Plantation, served as a representation for the three-dimensional maps, where strawberries were grown, as depicted in Figure 2a. The control room was situated within the Greenhouse Garden, where the operator remotely controlled the agricultural robots through a network. Figure 2c illustrates the operator's control screen during remote operation.

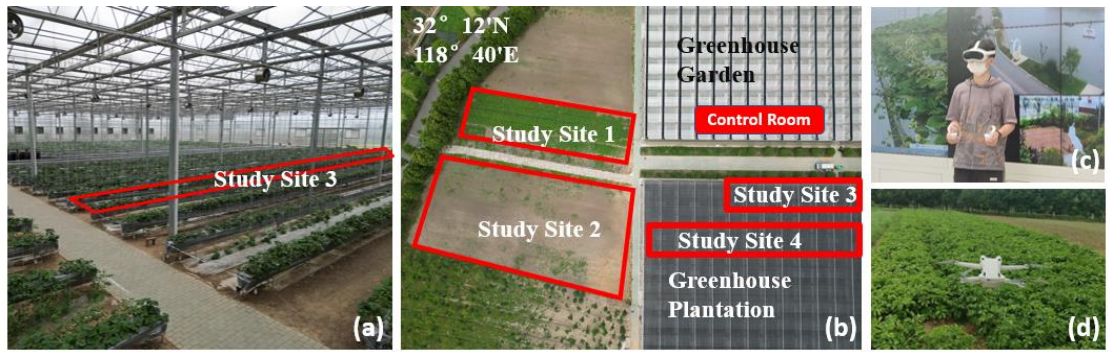

Figure 2 Study area. (a) Detailed location of study site 3 (b) General View of the Research site (c) Operator control within the control room (d) Drone Captured Footage

DJI® Mini 3 Pro drone (SZ DJI Technology Co., Shenzhen, China) is used to capture footage at 1080p resolution and 30fps. The drone flies at a height of 5m (Study site1,2) and 1.5m(Study site3) and adjusts the angle of view between 90° and 60°. The entire shooting process is controlled by a remote controller. Because of hand instability, the flight speed varies but is generally maintained between 1 m/s and 2 m/s. The drone follows a flight pattern resembling a series of connected "L" shapes. It moves first from a close point to a distant point, then performs a lateral movement to the left or right, and repeats this sequence.

## 2.3 Camera Calibration

Camera parameters are required during segmentation and subsequent reconstruction; therefore, camera calibration is performed first. This process is vital. It involves determining the camera's intrinsic parameters, which subsequently enable a more accurate reconstruction. For this purpose, a calibration board with a 7×10 checkerboard pattern was created. This pattern allows for the accurate determination of spatial relationships between different parts of the field of view. Twenty-five images with varying degrees of rotation and translation were captured to provide a robust and comprehensive calibration. These images were then used to compute the camera's intrinsic parameters.

The form of calculated camera intrinsic parameters is:

$$K = \begin{pmatrix} f_x & 0 & c_x \\ 0 & f_y & c_y \\ 0 & 0 & 1 \end{pmatrix} \quad (1)$$

The parameters $f_x$ and $f_y$ represent the focal length of the camera, measured in pixels. They correspond to the focal length in the horizontal and vertical directions of the image, respectively. The parameters $c_x$ and $c_y$ denote the coordinates of the principal point, also measured in pixels.

**2.4 Feature-based video segmentation**

For subsequent processing, it is necessary to split the captured video into individual images. If the videos are directly divided at regular intervals, the unstable speed of the drone will cause the overlap rate between images to be inconsistent. Areas with slow flight will have too much overlap, increasing computational cost. In fast-moving areas, reduced overlap complicates finding relationships between images and may hinder accurate camera pose estimation. In addressing the issue of variable drone speed, adopting fixed-distance shooting segmentation based on overlap can also be viable solutions (Gonçalves et al. 2021). However, when capturing irregular objects, the utilization of fixed-distance shooting and overlap-guided segmentation may yield suboptimal results. Certain complex textures, particularly in areas featuring intricate internal structures, such as congregations of concave surfaces, may be overlooked and not captured during the shooting process. However, utilizing feature points as a standard for overlap allows for segmentation based on the complex textures of these feature points. This approach is more robust compared to using the degree of image overlap as a standard. This method makes shooting more flexible and simplifies subsequent processing, making it more accessible for novices.

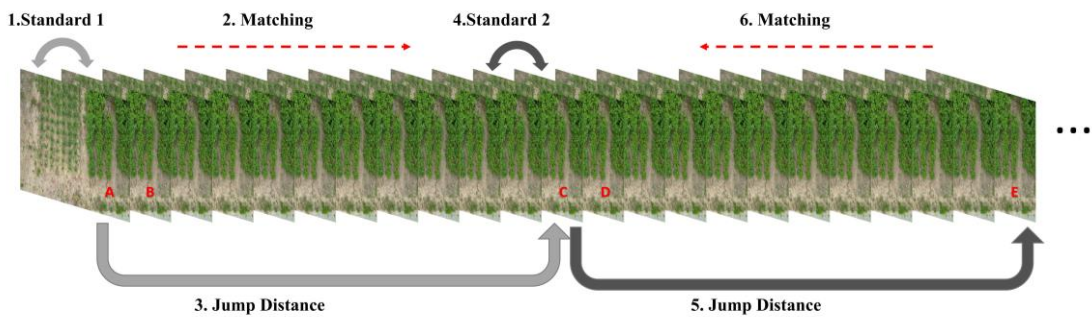

**Figure 3**　Flowchart of the Video Segmentation Algorithm

This method utilizes the SIFT algorithm to extract and match feature points between images. For each segmentation, an evaluation criterion was established based on the number of matched feature points between the saved frame and its adjacent frame. The subsequent images' feature points are continuously extracted until the number of matching points between the new image and the first image falls within the range of 50% to 33.3% according to the evaluation criteria. This approach ensures that all segmented images meet the computational requirements for subsequent calculations, while simultaneously reducing computational costs. To optimize the algorithm further, a strategy of determining the jump distance and conducting iterative searches is employed to enhance segmentation efficiency. After extracting feature points, adjacent images' feature points will be matched with each other, and only the successfully matched feature points are useful. Here, the number of successfully matched feature pairs is directly used as the criterion, rather than considering all the feature points in one image. The specific process is shown in Figure 3. Images A, C, and E will be retained, while other images will be discarded. The logarithms of matched feature points between images A and B and between images C and D will serve as criteria in their respective intervals. The workflow of the video segmentation algorithm is outlined in Algorithm 1.

---

**Algorithm 1** The working flow of the video segmentation algorithm
---
**Input:** videoPath: path to the input video, outputFolder: path to the output folder for saving frames, cameraParameters: camera intrinsic parameters according to Equation 1 .
**Output:** selectedFrames: list of selected frames.

1: $JD = 1$;　// Initialize the jump distance
2: **while** video has next frame **do**
3:　　Read($F_i$, $F_{i+JD}$);　//Read the currentFrame($F_i$) from the video
4:　　Save($F_i$);　//Save $F_i$ to the outputFolder
5:　　Extract($F_i$, $F_{i+JD}$)　//feature extraction by SIFT
6:　　$ev$=Match($F_i$, $F_{i+JD}$);
　　　　// Calculate the number of matching feature points between $F_i$ and the next frame as an evaluation criterion
7:　　**while** $ev/2 \leq mn \leq ev/3$ **do** //Filtering images

| | | |
|---|---|---|
| 8: | Extract($F_i$, $F_{i+JD}$) | //feature extraction by SIFT |
| 9: | $mn$ = Match($F_i$, $F_{i+JD}$); | |
| 10: | **if** ($ev/2 \leq mn$) **then** | //Jumping too close |
| 11: | $JD + +$; | |
| 12: | **else if**($mn \leq ev/3$) **then** | //Jumping too far |
| 13: | $JD = JD/2$; | |
| 14: | **end if** | |
| 15: | **end while** | |
| 16: | Save($F_{i+JD}$); | |
| 17: | $F_i = F_{i+JD}$; | //Update current frame |
| 18: **end while** | | |

## 2.5 Three-dimensional Point Cloud Reconstruction based on an Enhanced SfM

In reconstruction, OpenMVG is employed as the primary pipeline. OpenMVG is a multi-view based method for camera pose estimation and sparse reconstruction (Moulon et al., 2017). It directly processes multi-view images without needing additional image data. This reduces the reliance on GPS and makes it more suitable for complex agricultural environments. OpenMVG utilizes SIFT as its main feature matching technique, exhibiting strong robustness and outperforming the majority of current state-of-the-art algorithms. However, the performance of the SIFT algorithm is suboptimal in regions with poor texture. Due to the complexity of the agricultural domain, including the use of materials such as plastic films and greenhouses, as well as the presence of water surfaces, the uneven distribution of textures in the scene significantly reduces the applicability of OpenMVG in the agricultural field.

In this study, the advanced OpenMVG 2.0 version was utilized for incremental reconstruction tasks. Incremental reconstruction employs SIFT for feature extraction and matching. Features are extracted from all images, and for each feature point, a 128-dimensional descriptor (4×4×8) is formed. Feature matching is carried out so that each image is matched with every other image. In a set of n images, each image is matched n-1 times. This process consumes a considerable amount of time and computational resources. SIFT can more accurately link different frames together, but its matching quality may decrease for images with poor textures.

To improve feature matching in areas with poor texture, this study employs pre-trained

11 / 30

LoFTR weights. LoFTR is a Transformer-based feature extraction algorithm (Sun et al., 2021). It operates through a two-stage process: coarse-grained matching followed by fine-grained matching. The coarse-grained stage obtains the initial match locations, while the fine-grained stage refines these matches with more precision. The matching results are more accurate and robust after undergoing both coarse-grained and fine-grained steps. LoFTR's end-to-end learnable design allows it to obtain more matching points in regions with poor texture.

In this paper, LoFTR trained with Megadepth (2018) is employed. LoFTR has been integrated into OpenMVG and applied alongside SIFT to achieve more effective feature matching in regions with poor texture. Following SIFT processing, LoFTR will reprocess the feature points in the dataset. However, not all image pairs are subjected to LoFTR processing; we employ fixed window matching with a window size of 5, applying LoFTR only to image pairs within this window range. After obtaining the results from SIFT and LoFTR, the images are segmented into superpixels using the Simple Linear Iterative Clustering (SLIC) (Achanta et al., 2012) algorithm. The central formula of the SLIC algorithm calculates the distance between each pixel and the closest superpixel center. This distance computation enables the assignment of each pixel to its nearest superpixel center. The formula is represented as follows:

$$D = \sqrt{(L - L_c)^2 + (a - a_c)^2 + (b - b_c)^2 + \frac{(S - S_c)^2}{s^2}} \qquad (2)$$

Here, $D$ represents the distance between a pixel and the center of the corresponding superpixel. $L$, $a$, and $b$ denote the Lab color values of the pixel, while $L_c$, $a_c$, and $b_c$ represent the Lab color values of the superpixel center. Additionally, $S$ and $S_c$ correspond to the spatial positions of the pixel, and $s$ denotes the compactness coefficient, which is utilized to balance the influences of color and spatial distances. In this study, the predefined number of superpixels (segments) is set to 100, accompanied by a compactness coefficient of 10 and a Gaussian filter with a sigma value of 1.

Following superpixel segmentation, the feature point count within each superpixel and the average number of feature points contained within each superpixel per image are computed. As depicted in Figure 4d, distinct colors are employed to represent varying quantities of feature points, ranging from white to cyan, indicating increasing feature point abundance to scarcity. The standard is set as follows:

$$Standard = \frac{\sum_{i=1}^{n} FC_i}{2 \times n} \quad (3)$$

Here, $FC_i$ refers to the Feature Count of the $i$th superpixel. Feature points above this standard are considered to be in regions with rich texture, while those below the standard are classified as being in regions with poor texture. Only the LoFTR feature values in regions with poor texture are preserved. This ensures that feature points in regions with rich texture are from SIFT, while those in regions with poor texture are from both SIFT and LoFTR. However, LoFTR solely provides the XY coordinates of feature points and does not generate scale and orientation values. In this context, a nearest-neighbor search based on the squared Euclidean distance is utilized, as Equation 4.

$$D = (x - x_i)^2 + (y - y_i)^2 \quad (4)$$

The algorithm identifies the SIFT feature points closest to the feature points obtained through Loftr and replicates their scale and orientation values.

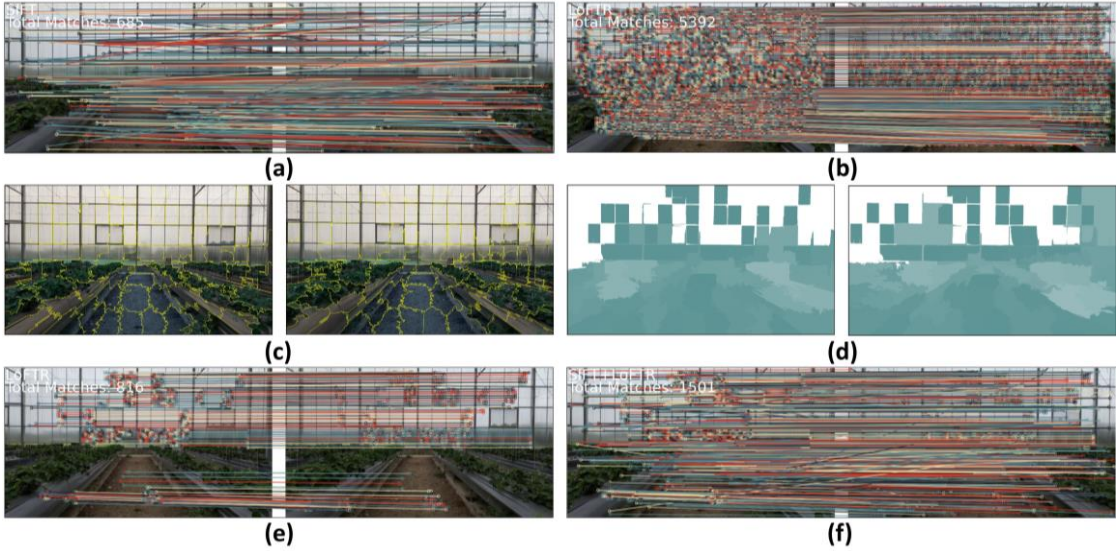

Figure 4 Overview of the proposed enhanced SfM (a) corresponding points by SIFT (b) corresponding points by LoFTR (c) superpixels by SLIC (d) texture analysis based on superpixels (e) Corresponding Points Retained by LoFTR in Poor Texture Regions (f) corresponding points by the proposed method.

The merged feature points are then filtered based on geometric consistency constraints, and incorrect matches from LoFTR and SIFT are removed. The algorithm starts from a single photograph and continually adds new images that have the most matching feature points with the current reconstruction. It calculates the spatial relationships between images based on feature points and then determines the corresponding camera poses. This involves solving the Perspective-n-Point (PnP) problem to estimate the new camera pose, and incrementally adding 3D points. During triangulation and camera positioning, Bundle Adjustment (BA) is used for optimization. While a significant amount of time is spent in BA, it effectively reduces errors and improves the accuracy of the reconstruction.

## 2.6 Digital Agricultural Map Construction Using MVS

After obtaining a sparse point cloud, OpenMVS is utilized to process it. Firstly, the DensifyPointCloud process is carried out. Preprocessing is performed on each of the images, addressing issues such as distortion and color correction. Depth estimation is then performed on each image, assigning depth values to individual pixels. The depth estimation results from all images are fused to create a complete, dense point cloud map, and outlier values are processed.

Subsequently, the dense point cloud map is converted into a 3D mesh through the ReconstructMesh step. During the 3D meshing process, a series of optimizations and improvements are carried out, including mesh simplification, denoising, smoothing, and hole filling, corresponding to the RefineMesh step. The processed 3D mesh is then assigned UV coordinates for texture mapping. Following this, a 2D texture atlas is generated, which is the TextureMesh step. To ensure the quality of the model, color blending is performed on texture atlases created from different images. Finally, the texture is assigned to the 3D mesh to generate an accurate textured 3D model.

## 2.7 Immersive Digital Map Viewing in VR

After obtaining the three-dimensional digital agricultural map, the map is used for immersive visualization display. Unity 3D is a powerful game development engine with strong capabilities in rendering and displaying 3D models, as well as processing and creating complex graphics. In addition, it supports cross-platform development on various operating systems, including Windows, Android, and Ubuntu. Unity is utilized for digital map and VR application development. The digital agricultural map is imported into Unity at a scale approximating reality, and through Unity's rendering and processing, it is presented with improved visualization.

In the display devices, ByteDance's PICO 4 VR device is utilized. The PICO 4 is a powerful VR device capable of showcasing a 4K+ resolution, 20.6 PPD, and a 105° wide field of view. Notably, the PICO 4 supports a high refresh rate of 90Hz. As a recent outstanding VR device, it has undergone significant optimizations in terms of weight and wearing comfort, making it more suitable for remote operation and extended usage. Furthermore, the PICO 4 is equipped with a Qualcomm Snapdragon XR2 processor and an Android system, allowing for direct application development without the need for computer-side streaming. Due to its excellent performance and comfort, the PICO 4 serves as an ideal VR device, assisting operators in prolonged remote operations.

In the development process, the project requires the use of the XR Interaction Toolkit package available in Unity. The XR Interaction Toolkit is a cross-platform tool for VR/AR interaction development, featuring a range of functionalities that support VR devices. It helps us obtain input and output information from VR devices, particularly regarding controller input and head pose input. To accommodate the PICO 4, it is necessary to import the PICO Unity Integration SDK provided by PICO's official source into the Unity project, ensuring compatibility with the PICO 4. Within the project, input from the left-hand controller is employed to control the movement of the virtual character's position. Operators can utilize the controller to move forward, backward, and sideways. Additionally, virtual reality captures changes in head orientation, which are utilized to control the rotation of the viewpoint.

**2.8 TCP/IP-based Closed-Loop Control for Remote Teleoperation**

The control end and the machine end are connected using the TCP/IP to establish communication. The machine-end computer serves as the server, while the PICO 4 functions as the client. The left-hand controller of the PICO 4 is used to control the movement of the character's viewpoint within the digital scene, while the input from the right-hand controller is directly collected and sent to the server end. TCP/IP communication is facilitated through Unity's .NET Framework. The StreamReader and StreamWriter classes are utilized to send and receive string messages. Upon receipt of control instructions from the PICO 4 at the robot end, the instructions are immediately forwarded to the server. The server's ROS system receives these instructions through Python, which in turn controls the movement of the robot.

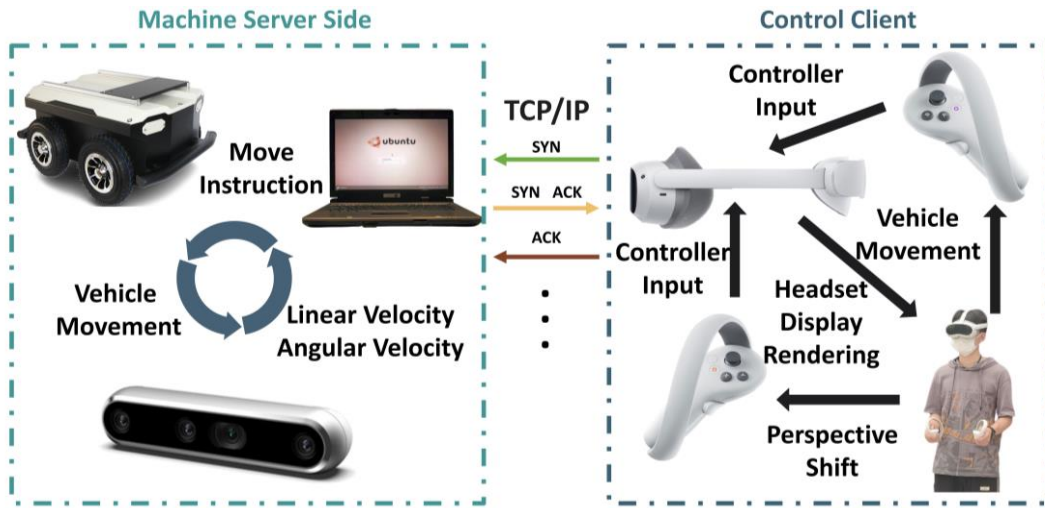

Figure 5 Framework for closed-loop control

Upon receiving the instructions, the ROS robot publishes a Twist message type to the cmd_vel topic. The Twist message type includes linear and angular vectors, representing linear velocity and angular velocity, respectively. cmd_vel is a standard ROS topic that communicates with the underlying drive node of the ROS robot. The entire process is seamlessly controlled through a publish/subscribe model that interfaces with hardware drivers, enabling ROS control.

In open-loop control scenarios, it becomes increasingly difficult to accurately predict the position of the mobile robot due to the accumulation of errors during motion. To address this issue, the RealSense D455 camera is employed as a visual odometry system to determine the robot's displacement and rotation angle. The RealSense D455 is a depth camera designed by Intel, capable of capturing both RGB and depth information. This camera is suitable for a shooting range of 0.6 to 6 meters, and its depth estimation error is controlled within 2% for objects within 4 meters.

In the project, the camera is mounted on the front of the mobile robot to calculate its movement by capturing depth changes on the ground in front of robot. Forward and backward motion primarily relies on the vertical changes in the depth image, while the rotation angle is determined by calculating the average distance on each side and the baseline distance. To reduce the impact of noise, the acquired depth information is processed through spatial and temporal filters. The filtered data is then used to compute average distances and rotation angles, enabling real-time estimation of the robot's position and pose. The robot's position and pose data are transmitted in real-time via TCP/IP to the virtual reality (VR) environment. The robot's position in VR is updated accordingly, providing the operator with a real-time understanding of the robot's status.

## 3. Results and Discussion
### 3.1 Impact of Video Segmentation Algorithms

To reduce a large amount of redundant data, an algorithm has been developed to segregate data based on feature points. In order to ascertain the effectiveness of the algorithm proposed, four sets of experiments were conducted, with each set operating at different flight speeds. The first set functioned at speeds between 0 and 1 m/s. The second set operated at speeds ranging from 0 to 2 m/s, while the third set varied from 0 to 4 m/s. For the fourth set, operations were conducted within a speed range of 0 to 4 m/s, but each flight trajectory involved two accelerations and two decelerations, one within a 0 to 1 m/s range and another within a 0 to 4 m/s range.

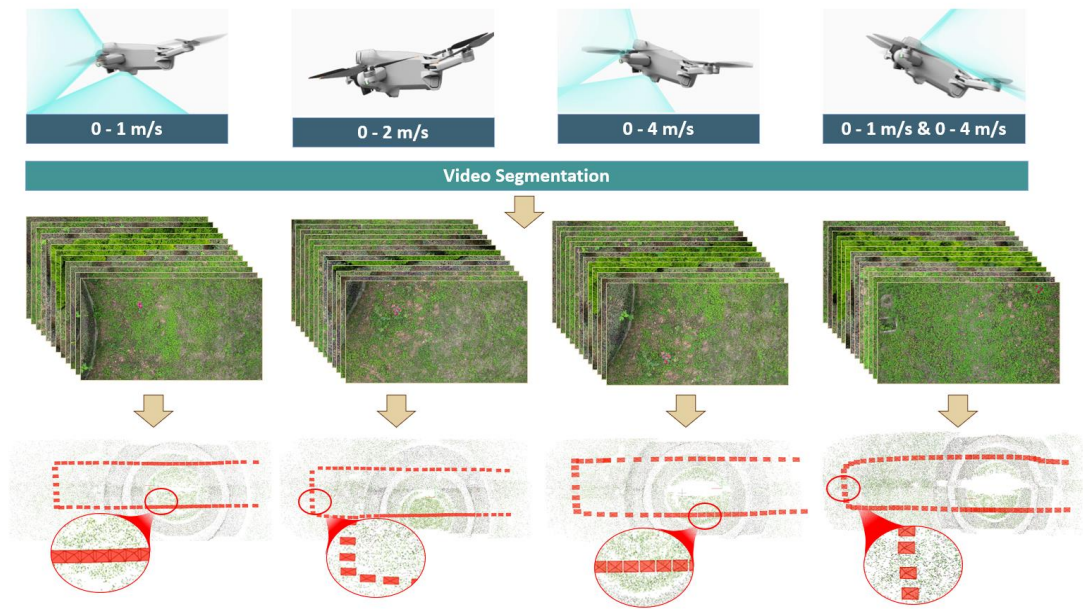

Figure 6 The video segmentation results are estimated by evaluating the information obtained at different flight speeds.

Due to the utilization of manual control during the experiment, our measurements might not have been absolutely precise, but they were sufficiently accurate to affirm the validity of the proposed algorithm. The results from these experiments were used to compute the camera pose. As can be seen in Figure 6, the number of images from the experiment was significantly reduced. Whether at straight line segments or corners, the overall distribution of images was relatively uniform. Especially at the corners, the drone's speed was reduced to zero before making lateral movements. This process, owing to alterations in drone speed and pauses, can result in a considerable volume of redundant imagery, thus escalating data volume. Nevertheless, our algorithm exhibited extraordinary effectiveness in this context.

Moreover, there was no loss of images throughout the derivation of camera pose; all images underwent camera pose derivation. The distribution of images is anchored on these feature points; in regions with rich textures, the images are more densely accumulated, and the quantity increases, thereby facilitating the acquisition of more comprehensive data. Conversely, in low texture regions, images are more scattered, and their quantity is reduced, thus lowering computational costs.

Table 1. Results of Video Segmentation for study site 1, 2, 3.

| Data Acquisition Area | Video duration | Number of original images | Number of processed images |
|---|---|---|---|
| Test Area 1 | 188.6s | 5655 | 417 |
| Test Area 2 | 191.0s | 5726 | 452 |
| Test Area 3 | 62.0s | 1860 | 141 |

Table 1 presents the results obtained after segmenting the video footage captured from the three study sites employed in our subsequent experiments. Shooting at a rate of 30 frames per second resulted in abundant data acquisition with a high volume of images. However, the process also generated a significant amount of redundant data. Following processing, a notable reduction in the volume of the resulting images was observed.

**3.2 The Performance of Reconstruction**

The experiments took place in three distinct testing areas: Test Area 1, Test Area 2, and Test Area 3. In each location, drones were utilized for filming purposes, following which our proposed video segmentation technique was applied to generate a respective count of 417, 452, and 141 images. Upon obtaining this data, a comparative study was conducted between the conventional SfM algorithm and our enhanced SfM method (SIFT+LoFTR).

In our point cloud reconstruction results, all input images had their camera poses successfully computed in multiple experiments. The SfM Scene Root Mean Square Error (RMSE) was utilized to estimate performance. RMSE computes the error between the reconstructed 3D points and the actual 3D points. It is the square root of the average of squared re-projection errors across all images and is frequently employed for assessing errors in the field of three-dimensional reconstruction. RMSE is commonly employed as a measure to assess the accuracy of 3D structure reconstruction or camera pose estimation.

$$RMSE = \sqrt{\frac{1}{N}\sum_{i=0}^{N}(\hat{x}_i - x_i)^2 + (\hat{y}_i - y_i)^2 + (\hat{z}_i - z_i)^2} \tag{5}$$

Here, $\hat{x}_i$, $\hat{y}_i$, and $\hat{z}_i$ are the predicted $x$, $y$, and $z$ coordinates of a point, respectively, whereas $x_i$, $y_i$, and $z_i$ represent the actual observed point locations.

To quantify the improvement, a statistical analysis was conducted on the "Number of Points" for both traditional and enhanced algorithms, thereby assessing the contribution of LOFTR to algorithmic enhancement. The point cloud is capable of representing scene information and can reflect the algorithm's performance in presenting the intricacies and complexity of the scene. Additionally, the determination of camera pose requires the participation of high-quality point clouds. To substantiate the validity of projected points during the 3D reconstruction process, "Points Validated" is employed as an evaluative measure. Verification is conducted for all projected points during the 3D reconstruction process. "Points Validated" represents the percentage of correctly matched point pairs within the total number of projected points.

Table 2. Comparison of Results between Enhanced SfM and Traditional SfM.

|  | Views | Method | RMSE | Number of Points produced by SfM | Points Validated |
|---|---|---|---|---|---|
| **Study site 1** | 417 | OpenMVG | 0.411572 | 630,342 | 0.736434 |
|  | 417 | OpenMVG+ LoFTR | 0.373865 | 1,222,507 | 0.968 |
| **Study site 2** | 452 | OpenMVG | 0.335612 | 210103 | 0.939655 |
|  | 452 | OpenMVG+ LoFTR | 0.270256 | 1,606,487 | 1 |

|  | 141 | OpenMVG | 0.434822 | 37358 | 0.928 |
| --- | --- | --- | --- | --- | --- |
| **Study site 3** | 141 | OpenMVG+ LoFTR | 0.368285 | 211771 | 0.967742 |

As illustrated in the table 2, the enhancements made to the SfM algorithm have resulted in an improvement in the RMSE. This indicates that the modified algorithm has improved robustness. Additionally, Table 2 shows a significant increase in the number of feature points identified by the improved algorithm. This observation is evident from Figure 7 a-f, where the application of the enhanced SfM method demonstrates a notably improved overall presentation. More information is obtained in areas with weak textures. A higher count and higher quality of feature points can contribute to a more precise estimation of camera pose, which is a critical element for successful SfM implementations. Also, the increase in feature points enables the algorithm to capture more intricate and rich information from the scene, thereby enhancing the algorithm's capability to manage complex environments and provide a more comprehensive representation of the 3D structure. The increment in "Points Validated" suggests that a higher percentage of identified points have been authenticated in the modified algorithm, further attesting to its effectiveness. The validation process discards inaccurately matched point pairs, ensuring that the remaining points make a positive contribution to the 3D reconstruction and camera pose estimation. Therefore, the increased validation rate in the improved algorithm supports the claim that it offers more accurate feature point detection.

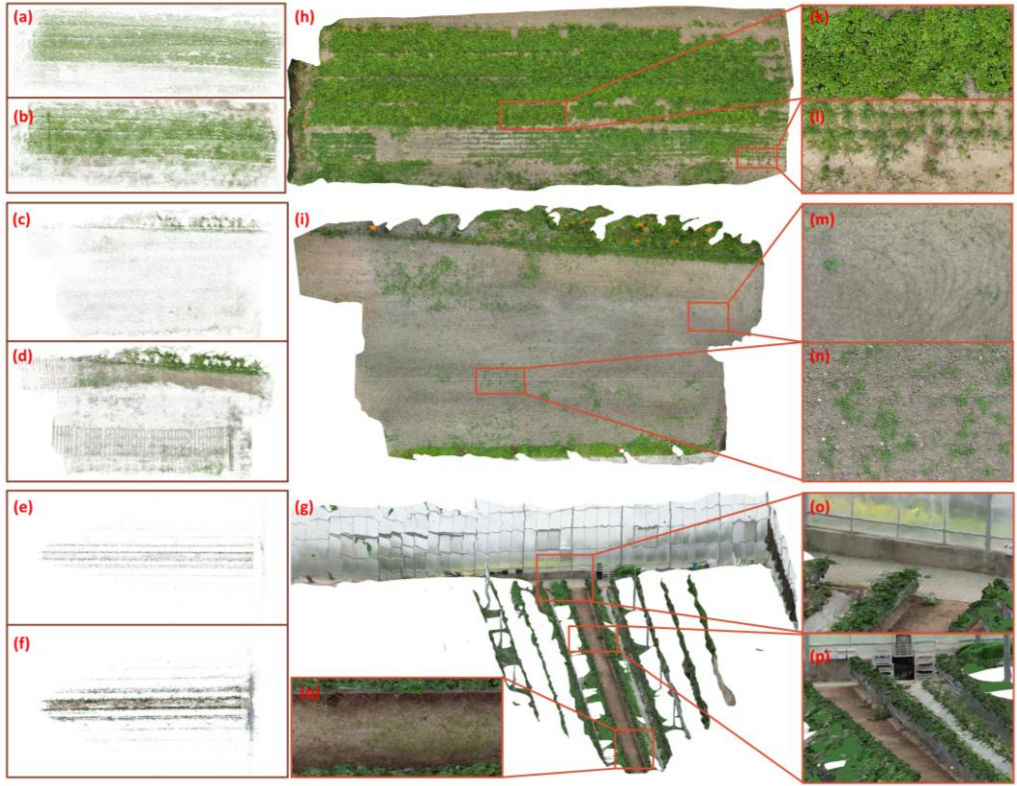

Figure 7 Reconstruction Results (a), (c), (e) Point Cloud Diagram of SfM for study site 1,2,3 (b), (d), (f). Point Cloud Diagram of the Proposed Enhanced SfM for study site 1,2,3 (h), (i), (g) Model Generated by the Proposed Reconstruction Method for study site 1,2,3 (k), (l), (m), (n), (o), (p), (q) Enlarged View of Model for study site 1,2,3.

After the application of the Proposed Enhanced SfM and MVS, the digital maps for three study sites are presented in Figure 7 h-g. Study site 1 and 2 depict a flat field, as evidenced by the enlarged maps, where even minute vegetation details have been accurately reconstructed. Figures 7k-l provide detailed views of ground-level vegetation, allowing individual leaves to be discerned. Figures 7m-n highlight the texture of bare cultivated land and the scattered weeds. Transitioning to Scene 3, which features a greenhouse environment, data acquisition was limited to three sides. Notably, figures 7o-q vividly demonstrate the successful reconstruction of corridor spaces and transparent glass surfaces.

### 3.3 Immersive viewing and control

The digital agriculture map was integrated into a Virtual Reality (VR) Android application using the Unity engine. Experiments were conducted in outdoor agricultural environments to validate the immersive display effects of the digital farm within the Virtual Reality (VR) setup. Once imported into Android, the scene undergoes rendering through Unity. The digital agriculture map not only replicates the appearance of an authentic agricultural landscape but can also present varying illumination effects under the rendering process.

Within the VR device, operators are fully immersed in a virtual farm setting. Figures 8c-e show the results of importing the map into Unity. Regrettably, the maps undergo a simplification process upon being imported into Unity, leading to a decrease in map quality. Despite this, the scene retains a quality that meets the intended standards. The maps are presented to operators, enhancing their immersive understanding of the real-world scenario. Moreover, the VR headset will present distinct visuals to each eye, providing operators with additional 3D information due to the disparity in perspectives between the two images as Figures 8 j-l. Traditional computer monitors only display the operational site's conditions, and the exterior of the monitor often distracts the operator. Even if it's possible to transmit information from multiple on-site cameras, thus gaining information from multiple viewpoints, the fixed nature of these viewpoints is not advantageous for outdoor environments. In contrast, our approach eliminates physical restrictions, allowing operators to "walk through walls" or "fly" within the virtual environment. By using the left-hand controller, operators can identify the optimal viewpoint, enabling movements and perspectives that would be impossible to achieve in real-world scenarios.

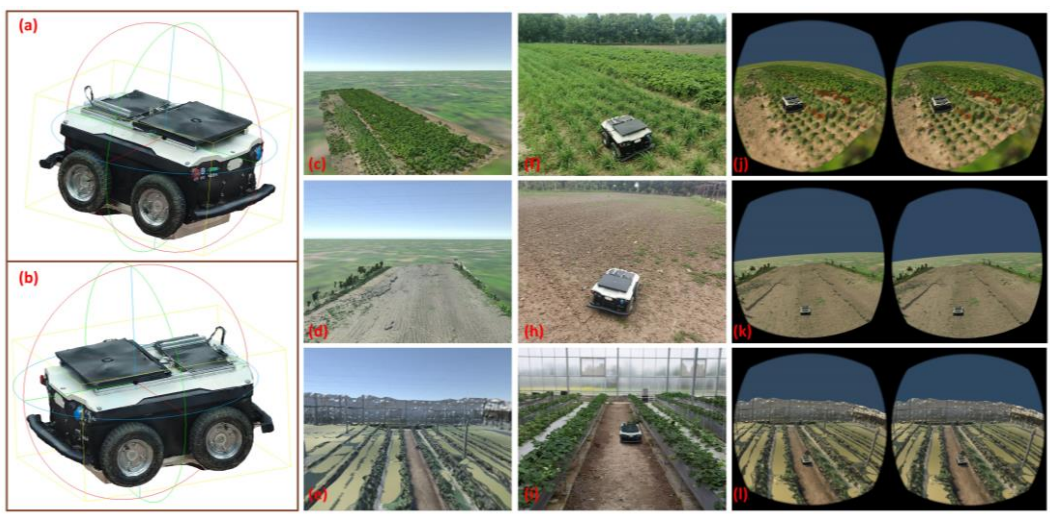

Figure 8 Remote Operation (a),(b) Model for
the mobile robot (c),(d),(e) Digital Agricultural Map Rendered in Unity (f),(h),(i) Machine movement screenshots in the real-world environment for study sites 1, 2, 3 (j), (k), (l) Machine movement screenshots in the Digital

Agricultural Map in VR for study sites 1,2,3.

To support remote operations, the mobile robot has also undergone a 3D reconstruction, and its model has been imported into the VR Android application, as illustrated in Figure 8a-b depicting the 3D model of the robot. Tests were conducted at three distinct study sites 1,2,3 and Figures 8f-l depict screenshots of vehicle movements in both the real-world environment and the VR environment. It is evident that VR enables the observation of optimal vehicle movement by allowing changes in perspective. This also validates the feasibility of the research framework.

**3.4 System assessment**

To assess the overall quality of system usage, a group of 20 volunteers was invited to participate in relevant testing. This group comprised 13 males and 7 females. Feedback from the volunteers was obtained through the utilization of a questionnaire. The questionnaire was used to get their feedback:

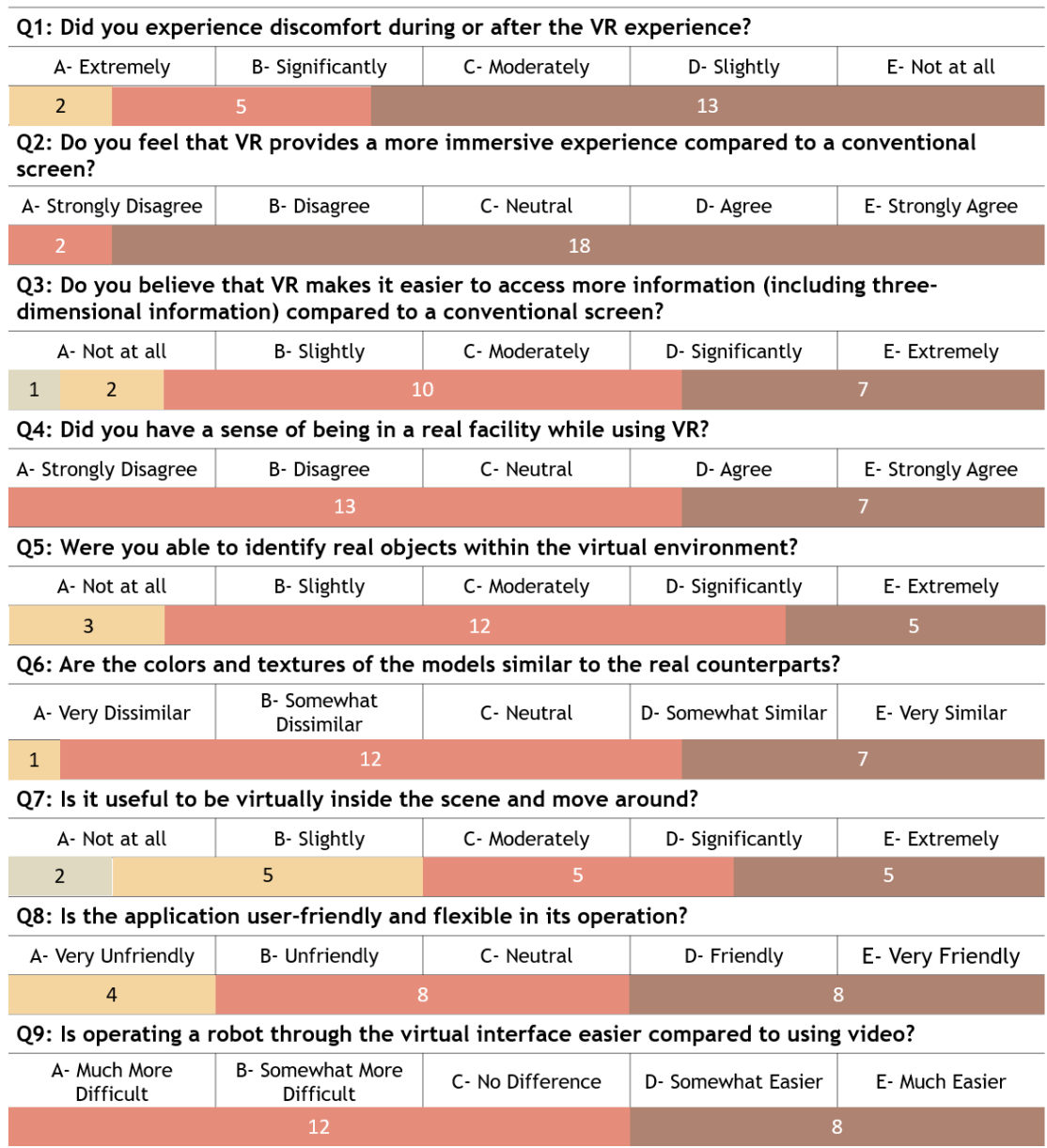

Figure 9 Questionnaire and results

Figure 9 displays the survey results. Different colors are used in the image to represent the number of responses for each option. Every third row (i.e., rows 3, 6, 9, etc.) represents the number of respondents for different options for each question. It reveals that 35% of the respondents reported experiencing a moderate to slight level of discomfort while using VR. The reasons for this discomfort can be attributed partly to VR technological limitations and partly to difficulties in adapting to the VR environment. If volunteers use the system repeatedly over an extended period and gradually adapt, it is expected that discomfort will decrease, similar to how motion sickness diminishes over time. The majority of volunteers highly acknowledge that the system can provide a greater sense of immersion and on-site information. Despite Unity's limitations reducing map quality, most operators find that object recognition is still effective.

Furthermore, the VR perspective offers a higher level of realistic detail, enhancing the immersive experience of control. Additionally, the freedom of unrestricted movement without physical constraints has provided volunteers with a level of convenience surpassing that of reality. 90% of the volunteers reported that changing the viewpoint freely in VR helped them. Furthermore, 100% of users found operating the robot through the virtual interface to be easier compared to using a computer screen. In summary, these findings indicate that the virtual interface is preferred for operating the robot, offering distinct advantages. While digital maps cannot be updated in real-time, they can be periodically refreshed through capturing and reconstruction to provide up-to-date information.

4. **Conclusions**

This study proposes a novel method for reconstructing agricultural digital maps and implementing a virtual reality remote control system. In the field of agricultural digital map reconstruction, a feature-based video segmentation technique is introduced. This technique demonstrates strong adaptability and enhances mapping accuracy. Particularly in structurally complex farms, this method proves to be highly practical. Fixed-distance drone flights controlled by GPS are impractical due to obstacles and diverse vegetation on farms, requiring manual flights. However, manual flights suffer from unstable speeds, resulting in excessive data collection in slower areas. This method effectively addresses this issue. Additionally, the SfM algorithm is optimized through integration of LoFTR and SIFT, enabling the generation of more reliable and comprehensive reconstruction results. This approach addresses the challenges posed by significant texture variations in farms, including water surfaces, plastic films, and various rough terrains, thus enhancing its suitability for farm use.

From a control perspective, a closed-loop control mechanism combining VR Android software with ROS robots is proposed. This method surpasses traditional screen-based approaches, enhancing operators' ability to acquire more comprehensive 3D scene information and ensuring immersive control effects. Investigations confirm and validate the effectiveness of the proposed method. However, the drawback of this approach is that the generated maps are static. Although the maps can be updated every few hours to account for sudden changes, integrating real-time cameras with video streaming would be a better solution. This would enable the capture of small-scale situations around the machine using cameras while providing operators with broader, extensive 3D farm information through our approach. Additionally, the system framework does not rely on robot hardware, facilitating experimentation on other machines. This flexibility provides a more efficient solution for remote robot control, enhancing agricultural management and facilitating tasks that may pose risks to human health.

## Acknowledgements

This work was supported by the National Natural Science Foundation of China (Project no. 32472010), the Jiangsu Agricultural Science and Technology Innovation Fund (JASTIF) (Grant no. CX(24)3027), and the Natural Science Foundation of Jiangsu Province (Grant no. BK20231478).

## References

Achanta, R., Shaji, A., Smith, K., Lucchi, A., Fua, P., & Süsstrunk, S. (2012). SLIC superpixels compared to state-of-the-art superpixel methods. *IEEE transactions on pattern analysis and machine intelligence*, *34*(11), 2274-2282.

Bazzano, F., Gentilini, F., Lamberti, F., Sanna, A., Paravati, G., Gatteschi, V., & Gaspardone, M. (2016). Immersive virtual reality-based simulation to support the design of natural human-robot interfaces for service robotic applications. In *Augmented Reality, Virtual Reality, and Computer Graphics: Third International Conference, AVR 2016, Lecce, Italy, June 15-18, 2016. Proceedings, Part I 3* (pp. 33-51). Springer International Publishing.

Benos, L., Bechar, A., & Bochtis, D. (2020). Safety and ergonomics in human-robot interactive agricultural operations. *Biosystems Engineering*, *200*, 55-72.

Berra, E. F., & Peppa, M. V. (2020, March). Advances and challenges of UAV SFM MVS photogrammetry and remote sensing: Short review. In *2020 IEEE Latin American GRSS & ISPRS Remote Sensing Conference (LAGIRS)* (pp. 533-538). IEEE.

Bian, F., Li, R., Zhao, L., Liu, Y., & Liang, P. (2018, August). Interface design of a human-robot interaction system for dual-manipulators teleoperation based on virtual reality. In *2018 IEEE International Conference on Information and Automation (ICIA)* (pp. 1361-1366). IEEE.

Ersahin, G., & Sedef, H. (2015). Wireless mobile robot control with tablet computer. *Procedia-Social and Behavioral Sciences*, *195*, 2874-2882.

Gonçalves, G., Gonçalves, D., Gómez-Gutiérrez, Á., Andriolo, U., & Pérez-Alvárez, J. A. (2021). 3D reconstruction of coastal cliffs from fixed-wing and multi-rotor uas: Impact of sfm-mvs processing parameters, image redundancy and acquisition geometry. *Remote Sensing*, *13*(6), 1222.

Green, M., Mann, D. D., & Hossain, E. (2021). Measurement of latency during real-time wireless video transmission for remote supervision of autonomous agricultural machines. *Computers and Electronics in Agriculture*, *190*, 106475.

Hoshi, S., Ito, K., & Aoki, T. (2022, October). Accurate and Robust Image Correspondence for Structure-From-Motion and its Application to Multi-View Stereo. In *2022 IEEE International Conference on Image Processing (ICIP)* (pp. 2626-2630). IEEE.

Ji, W., Pan, Y., Xu, B., & Wang, J. (2022). A real-time apple targets detection method for picking robot based on ShufflenetV2-YOLOX. *Agriculture*, *12*(6), 856.

Jin, Y., Liu, J., Xu, Z., Yuan, S., Li, P., & Wang, J. (2021). Development status and trend of agricultural robot technology. *International Journal of Agricultural and Biological Engineering*, *14*(4), 1-19.

Krupke, D., Starke, S., Einig, L., Zhang, J., & Steinicke, F. (2018). Prototyping of immersive HRI scenarios. In *Human-Centric Robotics: Proceedings of CLAWAR*

*2017: 20th International Conference on Climbing and Walking Robots and the Support Technologies for Mobile Machines* (pp. 537-544).

Li, Z., & Snavely, N. (2018). Megadepth: Learning single-view depth prediction from internet photos. In *Proceedings of the IEEE conference on computer vision and pattern recognition* (pp. 2041-2050).

Lipton, J. I., Fay, A. J., & Rus, D. (2017). Baxter's homunculus: Virtual reality spaces for teleoperation in manufacturing. *IEEE Robotics and Automation Letters*, *3*(1), 179-186.

Liu, Y., Novotny, G., Smirnov, N., Morales-Alvarez, W., & Olaverri-Monreal, C. (2020, October). Mobile delivery robots: Mixed reality-based simulation relying on ros and unity 3D. In *2020 IEEE Intelligent Vehicles Symposium (IV)* (pp. 15-20). IEEE.

Lv, Q., Lin, H., Wang, G., Wei, H., & Wang, Y. (2017, May). ORB-SLAM-based tracing and 3D reconstruction for robot using Kinect 2.0. In *2017 29th Chinese control and decision conference (CCDC)* (pp. 3319-3324). IEEE.

Meinen, B. U., & Robinson, D. T. (2020). Mapping erosion and deposition in an agricultural landscape: Optimization of UAV image acquisition schemes for SfM-MVS. *Remote Sensing of Environment*, *239*, 111666.

Mizuchi, Y., & Inamura, T. (2017, December). Cloud-based multimodal human-robot interaction simulator utilizing ros and unity frameworks. In *2017 IEEE/SICE International Symposium on System Integration (SII)* (pp. 948-955). IEEE.

Moulon, P., Monasse, P., Perrot, R., & Marlet, R. (2017). Openmvg: Open multiple view geometry. In *Reproducible Research in Pattern Recognition: First International Workshop, RRPR 2016, Cancún, Mexico, December 4, 2016, Revised Selected Papers 1* (pp. 60-74). Springer International Publishing.

Murakami, N., Ito, A., Will, J. D., Steffen, M., Inoue, K., Kita, K., & Miyaura, S. (2008). Development of a teleoperation system for agricultural vehicles. *Computers and Electronics in Agriculture*, *63*(1), 81-88.

Opiyo, S., Zhou, J., Mwangi, E., Kai, W., & Sunusi, I. (2021). A review on teleoperation of mobile ground robots: Architecture and situation awareness. *International Journal of Control, Automation and Systems*, *19*, 1384-1407.

Pérez, L., Diez, E., Usamentiaga, R., & García, D. F. (2019). Industrial robot control and operator training using virtual reality interfaces. *Computers in Industry*, *109*, 114-120.

Qin, Y., Su, H., & Wang, X. (2022). From one hand to multiple hands: Imitation learning for dexterous manipulation from single-camera teleoperation. *IEEE Robotics and Automation Letters*, *7*(4), 10873-10881.

Roldán, J. J., Peña-Tapia, E., Garzón-Ramos, D., de León, J., Garzón, M., del Cerro, J., & Barrientos, A. (2019). Multi-robot systems, virtual reality and ROS: developing a new generation of operator interfaces. *Robot Operating System (ROS) The Complete Reference (Volume 3)*, 29-64.

Slob, N., Hurst, W., Van de Zedde, R., & Tekinerdogan, B. (2023). Virtual reality-based digital twins for greenhouses: A focus on human interaction. *Computers and Electronics in Agriculture*, *208*, 107815.

Stotko, P., Krumpen, S., Schwarz, M., Lenz, C., Behnke, S., Klein, R., & Weinmann, M. (2019, November). A VR system for immersive teleoperation and live exploration with a mobile robot. In *2019 IEEE/RSJ International Conference on Intelligent Robots and Systems (IROS)* (pp. 3630-3637). IEEE.

Su, Y., Chen, X., Zhou, T., Pretty, C., & Chase, G. (2022). Mixed reality-integrated 3D/2D vision mapping for intuitive teleoperation of mobile manipulator. *Robotics and Computer-Integrated Manufacturing*, *77*, 102332.

Sun, D., Liao, Q., Kiselev, A., Stoyanov, T., & Loutfi, A. (2020). Shared mixed reality-bilateral telerobotic system. *Robotics and Autonomous Systems*, *134*, 103648.

Sun, G., & Wang, X. (2019). Three-dimensional point cloud reconstruction and morphology measurement method for greenhouse plants based on the kinect sensor self-calibration. *Agronomy*, *9*(10), 596.

Sun, J., Shen, Z., Wang, Y., Bao, H., & Zhou, X. (2021). LoFTR: Detector-free local feature matching with transformers. In *Proceedings of the IEEE/CVF conference on computer vision and pattern recognition* (pp. 8922-8931).

Sun, W., Iwataki, S., Komatsu, R., Fujii, H., Yamashita, A., & Asama, H. (2016, December). Simultaneous tele-visualization of construction machine and environment using body mounted cameras. In *2016 IEEE International Conference on Robotics and Biomimetics (ROBIO)* (pp. 382-387). IEEE.

Thakur, A., Venu, S., & Gurusamy, M. (2023). An extensive review on agricultural robots with a focus on their perception systems. *Computers and Electronics in Agriculture*, *212*, 108146.

Valiton, A., Baez, H., Harrison, N., Roy, J., & Li, Z. (2021, May). Active Telepresence Assistance for Supervisory Control: A User Study with a Multi-Camera Tele-Nursing Robot. In *2021 IEEE International Conference on Robotics and Automation (ICRA)* (pp. 3722-3727). IEEE.

Yang, X., & Jiang, G. (2021). A practical 3D reconstruction method for weak texture scenes. *Remote Sensing*, *13*(16), 3103.

Zhang, W., Zhu, Q., Zhang, T., Liu, H., & Mu, G. (2024). Design and control of a side dense transplanting machine for sweet potato seedlings on mulch film. *Computers and Electronics in Agriculture*, *224*, 109193.

Zhang, Y., Chen, C., Wu, Q., Lu, Q., Zhang, S., Zhang, G., & Yang, Y. (2018). A Kinect-based approach for 3D pavement surface reconstruction and cracking recognition. *IEEE Transactions on Intelligent Transportation Systems*, *19*(12), 3935-3946.